\documentstyle[pre,preprint,aps,epsf]{revtex}
\draft
\tighten

\begin{document}

\title{Quantum discreteness and fundamental 1/f noise in tunnel 
junctions, nano-composites and other many-electron systems} 
\author{Yu.E.Kuzovlev, Yu.V.Medvedev $^a $ , A.M.Grishin $^b $ \\
$^a $ A.A.Galkin DonPTI NASU, Donetsk, Ukraine \\
$^b $ Royal Institute of Technology, Stockholm, Sweden } 
\date{}
\maketitle

\begin{abstract}
It is shown, with citing tunnel junction as an example, that mutual 
interplay of the electron quantum transfers in a conducting system can 
be the fast mechanism for generation fundamental low-frequency flicker 
conductance fluctuations (1/f noise) without composing Lorentzians. 
This effect is lost in a theory which neglects the actual discreteness 
of electron energy levels. The analytical estimates of fluctuations of 
tunnel conductance are obtained, and the strong 1/f-noise sensibility 
to the discreteness as observed in nano-composites is explained both 
qualitatively and quantitatively. 
\end{abstract}

\section{Introduction}

1. The low-frequency flicker noise (1/f-noise) known in a large
variety of systems is actual problem of applied and theoretical 
physics [1-5]. This noise manifests itself as fluctuations in
rates of relaxation and transport processes. In electronics thats
are conductance fluctuations in practice often representing most bad
type of noise. As a rule, these fluctuations are much higher
sensitive to a structure of materials and external influences than 
conductivity itself and hence may bring a delicate information about
mechanisms of charge transport. But the problem is that there is no
adequate comprehension of the problem. 
  
For many years theory tries to reduce flicker noise to large 
temporal scales (the concepts of hierarchycal kinetics [5] and 
self-organized criticality [21] involve also large spatial scales). 
Usually, flicker noise is thought be result of slow thermally 
activated fluctuations in structural disorder or in occupancy of 
electron traps, etc. [1-3,5-15,29,37]. Spectrum 1/f is composed 
by Lorentzians corresponding to "fluctuators" with different 
activation energies, under assumption that sufficiently broad 
and uniform distribution of the energies takes place [1,5-15,37]. 
But in fact either qualitative origin of fluctuators remains unclear
or quantitative treatment of experiments remains made at a stretch.
Many important facts in no way could be kept within this theory,
for instance, 1/f mobility fluctuations of carriers in clear 
intrinsic semiconductors [2,10], or 1/f-noise in liquid metals [4].

At present, the noise in systems with bad (narrow-band, hopping, 
tunnel, percolative) conductivity is under common attention, namely, 
in strongly doped or defected or amorphous semiconductors [5,7,11-14]
and metals [8,42], in oxides [6,9], in manganites with colossal
magneto-resistance [16-20], etc. Since in such systems long-range
Coulombian forces can be significant, the charge redistributions
were suggested to the role of slow fluctuators [5,11-15]. But again
it is hard to explain the absence of saturation of real 1/f spectra
at low frequencies [7,15]. Perhaply, it was the true prophecy [22]
that the game "1/f from Lorentzians"  may turn to a disaster for
1/f-noise theory.
 
2. With no doubts of the importance of thermally activated "slow" 
fluctuations, it would be reasonable to separate them from the 
fundamental 1/f-noise whose direct source is nothing but the same 
"fast" kinetic events (interactions and collisions of 
particles and quanta) which produce the electric resistivity 
itself [4,23-27]. 

Let us recall that generally the phenomenon of 
relaxation and irreversibility (in particular resistivity) in one 
or another dynamical system is equivalent to its ability to 
constantly forget its own history. Most easy the oblivion spreads 
to the number of kinetic events which took place in the past 
(for instance, more than a few inelastic free-path times ago). 
Hence, the system loses both stimulus and means for keeping some 
definite ''number (or probability) of events per unit time'' (more 
accurately, only the ratio of numbers of time-reversed events is 
under thermodynamical control, but not their sum or difference) 
[23,26,27]. Therefore, the fluctuations in ''thickness'' (the 
term from [28]) of kinetic events and correspondingly in the rates 
of relaxation and transport do not cause some back compensating 
reaction, hence, they themselves do not relax and have no upper 
time scale. 

Such statistical behaviour results just in scale-invariant
non-saturating spectrum 1/f (for details see [4,27]). Thus, the 
extremes converge: long-living correlations associated with 
1/f spectrum reflect not a long memory but, in opposite, 
indifference with respect to the past (it is usefull, as
recommended by Krylov [8], to get rid of the prejudice that any
statistical correlation testifies some actual causal correlation). 

The kinetic theory throws out this child at the same time as 
turbid water if postulating quite certain ''probabilities per unit
 time'' (i.e. collision integrals etc.). But if the path from
statistical mechanics of a gas to gas kinetics is overcomed without
such an ansatz [24,27] then it leads to 1/f fluctuations of
diffusivities and mobilities of molecules. The origin of these
fluctuations is not long relaxation times (surely absent here)
but merely indifference of the system with respect to impact
parameters of previous collisions (and thus to their effective 
cross-section). The flicker fluctuations in dissipation and in
light scattering in quartz also can be explained as statistical
property of phonon kinetics [25]. The latter does not reduce to
a certain marginal collision integrals, because phononic kinetic
events are strongly entangled in space and time and parametrically
interplay one with another. 

3. Hence, the general principle of ''1/f-noise from lossing 
memory'' realizes by different concrete ways. In this work 
(representing development of the preliminary study [46]) we argue 
that in many-electron systems it may realize like in phononic systems 
by means of mutual entangling kinetic events, i.e. electron 
transfers and jumps. We will see that flicker conductance 
fluctuations naturally come to light if a theory takes into account 
factual finite duration of kinetic events and besides real
discreteness of electron energy states. 
  
Because of its finite duration any one-electron transfer becomes 
the part of a complex many-particle process. If cutting off this part 
across bosonic lines one gets that quantum amplitude of given 
transfer evolves under influence by effective time-varying fields 
representing other components of the process. For instance, 
currently realizing electron jumps through a tunnel junction,
as well as thermal charge motion in its sides, induce 
fast voltage fluctuations at the junction. In its turn, this 
voltage noise causes random phase changes in the increments 
of quantum amplitude of the maturing transfer. 
Such a picture with mutual influence of one-electron events 
mediated by many soft photons was considered in the theory 
of Coulomb blockade and low-temperature anomalies of 
current-voltage cvharacteristics (CVC) of small tunnel
junctions [31-33]. The mathematically close interpretation
of many-phonon processes was used in the theory 
of mobility of strongly coupled polarons [34].  

Obviously, this picture must foresee not only renormalization 
of transport characteristics but in addition their specific 
fluctuations. If quantum amplitude of any electron transfer
becomes a functional of enviromental noise (electric, elastic, 
exchange fields, etc.) then corresponding quantum probabilities 
become random values, and eventually the conductance as well. 
As far as we know, previously this phenomenon
was not under consideration. It must be well expressed if
correlation time of the noise is small as compared 
with typical duration (expectation time) of electron jumps. We 
will demonstrate that such a situation is easily realizable, and 
the result of fast enviromental noise, regardless of its 
concrete mechanisms and statistics, is just flicker scaleless 
fluctuations of transfer rates (probabilities) and conductance 
(therefore, flicker noise what accompanies hopping conductivity
does not need in slowness of charge redistributions, instead
such thats are sufficient which are as fast as diffusion of
carriers). 

As the experiments prompt, in relative units conductance 
fluctuations are inversely proportional to a number of movable 
carriers in a noisy region, that is to its volume. From here the 
principal statement does follow: flicker noise can not be cathed  
by a theory which underestimates role of the discreteness and 
too hurries to turn sums over energy states into integrals. 

4. For visuality, we will concentrate at ''ideal'' tunnel junction.
Previously various aspects of many-particle processes and the 
discreteness in tunnel junctions were touched (see e.g. [35,36] 
and refs. therein), but with no respect to the effect of our
interest. This effect, i.e. conductance fluctuations caused by fast
thermal noise, principally differs of the low-temperature
conductance fluctuations in tunnel micro-junctions due to
quasi-elastic electron scattering by static structural disorder.
[44,45]. The 1/f-noise observed in ''material'' junctions [37,41]
usually is attributed to structural fluctuators, so called
two-level systems. The possibilities of this theory (developing
general approach [1]) were analysed in [30]. This theory
surely explains Lorentzian contributions to the whole noise
but avoids an interpretation of the ''residual'' 1/f component
observed in [37]. Unlikely one might categorically say that
this component (and then almost all high-temperature noise
possessing smooth 1/f spectrum) can not have another origin.
Such a suspicion is supported by the experiment [41] as
discussed below. 

\section{Time scales of tunnelling} 

1. If a voltage $U<T/e$ ($T$ is temperature) is applied to
a tunnel junction then the mean charge transported during
time $\Delta t$ and corresponding conductance can be expressed as
\begin{equation}
\Delta Q=e\cdot \frac{Ue}{\delta E}\cdot 
\frac{\Delta t}{\tau _t} \,,\,
\,G=\frac{\Delta Q}{U\Delta t}=e^2\nu \gamma \,,\,\,
\nu =\frac 1{\delta E}\,,\,\,\gamma=\frac 1{\tau _t}
\end{equation}
Here $\delta E$ is mean spacing of electron energy levels, 
thus $\nu $ is density of states; $Ue/\delta E$ is the number of 
''active'' levels effectively contributing to electron transport 
between outward leads; $\tau _t $ is mean transmission time 
required for one electron jump from a given level at one side to 
wherever at opposite side, in other words, this is the time of 
accumulation of quantum jump probability to a value $\sim 1$  
 (though quantum jumps are momentary, the moment is
unpredictable and may be expected up to $\tau _t$ or longer); 
 $\gamma $ plays the role of mean jump probability per
unit time (thus $e/\tau _t $ is the mean current per active level).

Any real junction has a finite capacity $C$ and thus
characteristic time $\tau _{c} \approx RC\equiv C/G $ . 
Its physical sense is the relaxation time of charge
captured by $C $ and the correlation time of thermal voltage
fluctuations at the junction. Compare the defined time scales
and let be convinced of that tunneling is very long event, that is  
\begin{equation}
\frac{\tau _t}{\tau _c}= \frac{e^2\nu }C=
\frac{E_c}{\delta E}>>1
\end{equation}
even if Coulombian effects are weak in the trivial 
sense $E_c<<T$ . For certainty, consider flat metallic
electrodes with thickness $w $ , correspondingly flat
barrier with thickness $d$ and typical
dielectricity $\epsilon \sim 20 $ . Then taking from textbooks 
standard estimates for capacity and Fermi energy and velocity
one obtains  
\[
\frac {e^2\nu }{C} \approx \frac{e^2}{\hbar v_F}\cdot 
\frac{4\pi dw}{\epsilon a^2}
\approx \frac{dw}{a^2} \,\,,
\]
where $a$ is atomic size, $a \sim 3\cdot 10^{-8}\, cm $ .

This expression clearly shows that inequality (2) is always 
satisfied. Consequently, while tunneling an electron has 
time to virtually feel many variations of the inter-side 
voltage noise, $u(t)$ , produced by many other jumps in 
both directions. At one-electron language, this means that 
quantum-mechanical transfer probabilities behaves as random 
processes. From the point of view of rigorous field theory of 
many particles, the description of corresponding excess 
fluctuations of transport current would need in at least 
four-particle Green functions [4]. Since the necessary formal 
technique is still absent, we will try to understand 
the essence of the matter by means of archaic tunnel Hamiltonian 
method and quantum-mechanical non-stationary perturbation theory. 
We confine ourselves by the simplest variant: the sides are 
identical, all tunnel matrix elements  $g_{kq}\approx g $ 
are approximately equal and transparency is small. 

2. Let us recall the underlying reasons of standard
tunnel Hamiltonian scheme (for a time forgetting voltage 
noise). Let  $p_{kq}(\Delta t,U) $  be the probability 
of electron transfer from a state "k" at left side to 
a state "q" at right side during an observation  
time $\Delta t $ (or probability of the time-reversed event).  
The probability of electron jump from left level "k" 
to wherever at right-hand side is given by sum of these  
elementary probabilities, and jump duration 
is the inverse jump probability per unit time: 
\[
\gamma =\tau _t^{-1}=p_k(\Delta t,U)/\Delta t \,,\,
\, p_k(\Delta t,U) \equiv \sum_{q}p_{kq}(\Delta t,U)  
\]
(by above assumptions, these values are weakly dependent on "k"). 

Further, the standard scheme recommends (as usually when 
constructing kinetics) to resort to the Fermi ''golden rule'',  
\[
p_{kq}(\Delta t,U) \rightarrow 2\pi g_{kq}^2
\delta (E_{kq})\Delta t/\hbar \,,\,
\,\,\,E_{kq}\equiv E_q^{+}-E_k+eU 
\]
(plus relates to right-hand side). This recipe ensures the
time-linear increase of the jump probability and thus the 
existence of certain rate of jumps,  
\begin{equation}
\gamma =\tau _t^{-1}\approx 2\pi g^2\nu /\hbar
\end{equation}
The using golden rule (in other words, continuous spectrum 
approximation) implies that the duration of watching on 
the evolution of quantum amplitudes, $\Delta t $ , what is  
sufficient for adequate evaluation of jump probabilities 
is in the frames  $\hbar/T <<\Delta t <<\tau_{g} $ . 
Here the characteristic time is 
introduced, $\tau_{g} =2\pi\hbar /\delta E $ , 
involved by discreteness of energy spectrum. 

Clearly, if one wants to account for the influence of voltage 
noise $u(t) $ then the adequate watching time should be much 
greater than its correlation time, and in view of (2) it 
is desirable to continue the watching time up to a value 
of order of factual duration of jumps. Hence, the standard 
scheme needs in the condition  $\tau_g >\tau_t $ . 
But the Eqs.1 and 3 lead to the relation 
\begin{equation}
\frac{\tau _t}{\tau _g}= 
\frac R{R_0}\approx \left( 
\frac{\delta E}{2\pi g}\right)^2 \,,\,
\,\,R_0\equiv \frac{2\pi \hbar }{e^2}\,,
\end{equation}
which shows that in a bad transparent junction just the 
opposite situation takes place, 
\begin{equation}
\tau _t/\tau _g>>1
\end{equation}

Thus it comes out that the golden rule is inapplicable, and 
one runs into the problem: ''probabilities per unit time'' 
remain uncertain. 

We may take in mind the case of well expressed discreteness (5), 
since this is principally most interesting case and besides 
at $R<R_0 $ compatibility of the tunnel Hamiltonian method  
and perturbation theory could be under question [38] (nevetheless, 
junctions with  $R<R _0 $ also will be described if thats can 
be represented as parallel sum of autonomous high-resistance
junctions). Under condition (5), the perturbation theory
is undoubtly applicable even in presence of voltage noise. 

\section{Fluctuations of probabilities}

1. Consider quantum transfers influenced by time-varying potential 
difference, $u(t)$ , between initial and final states. 
According to modern theory of quantum chaos, stochastic behaviour 
is typical for quantum systems regardless of energy discreteness    
[39,40]. Therefore, we will treat  $u(t)$ like classical random 
process (though remembering that in rigorous theory  $u(t)$ is 
an operator entangled with particle operators). 
At $\Delta t \sim \tau _t $ , the perturbation theory is 
sufficient to approximately solve the Shroedinger equation for 
wave function of tunneling particle. The result reads as  
\begin{equation}
p_{kq}\approx \left| A_{kq}\right| ^2\,,\,
\,A_{kq}\equiv \frac{g_{kq}} \hbar 
\int_0^{\Delta t}\exp (iE_{kq}t/\hbar )\,Z(t)dt\,,\,
\,Z(t)=\exp [i\varphi (t)] \,, 
\end{equation}
where the diffusively accumulating random phase is introduced, 
\[
\varphi (t)=\frac e\hbar \int_0^tu(t^{\prime })dt^{\prime } 
\]
At $u(t)=0$ these formulas reduce to usual ones, 
otherwise they describe chaotic parametrical excitation 
or damping of probabilities by the random phase.   

Let us introduce phase correlation function, corresponding 
coherence time of quantum amplitudes and "coherence region" 
by formulas 
\begin{equation}
K(t_1-t_2)=\left\langle Z(t_1)Z^{*}(t_2)\right\rangle \,,
\,\,\tau _{coh}=\int_0^\infty \left| K(\tau )\right| d\tau \,,
\,\,\Delta E=2\pi \hbar /\tau _{coh} \,,
\end{equation}
where angle brackets denote the averaging with respect to $u(t)$ . 
Generally, because of multiplicative structure of  $Z(t)$ 
calculations of even simplest statistical characteristics of the 
solution require a vast statistical information about  $u(t)$ . 
But if the coherence time is much smaller than observation time 
then factor $Z(t)$ under integral in (6) acts 
as complex fast (''white'') noise. Consequently, the transfer 
amplitudes $A_{kq} $ behave as (complex) Brownian walks, and 
the minimal information presented by characteristics (7) 
is already sufficient. 

2. Discuss briefly the coherence time, assuming for simplicity  
that shunting influence by external circuit on  $\tau _c $ 
is negligible. Note that $K(t)$ represents characteristic 
function of the phase, in the sense of probability theory. 
This function can be easy written at $E_{c}<<T $ if one treats 
voltage noise as Gaussian random process. Corresponding 
calculations give  
\[
\tau _{coh}\sim (\hbar /e)(C/T)^{1/2} <<\tau _g 
\]
(perhaply, that is minimally possible value of $\tau_{coh} $ ). 
At $E_{c} \sim T $ , charge quantization is essential and it is 
better to treat $u(t) $ like triple-valued random process, 
 $u(t)=0,\pm e/C $ . Corresponding analysis finishes
with simple estimate 
\[
\tau_{coh} \sim \tau_c 
\]
In this case coherence time may be comparable with  $\tau_g $ . 

Hence, there are all grounds to consider $Z(t)$ as white 
noise and consequently the amplitudes as Brownian walks. This 
allows to write   
\begin{equation}
\left\langle p_{kq}^2\right\rangle =
\left\langle \left| A_{kq}\right| ^4\right\rangle 
\approx 2\left\langle \left| A_{kq}\right| ^2
\right\rangle ^2\,=2\left\langle p_{kq}\right\rangle ^2
\,,\,\,\left\langle p_{kq},p_{kq} \right\rangle 
\approx \left\langle p_{kq}\right\rangle ^2
\end{equation}
We use Malakhov cumulant brackets, 
\[
\left\langle x,y\right\rangle \equiv \left\langle xy\right\rangle
-\left\langle x\right\rangle \left\langle y\right\rangle 
\]
At first, we see that after a time longer than coherence time 
the quantum amplitudes and probabilities acquire 100-percent 
uncertainty. Secondly, the mean probabilities grow linearly 
with time: 
\begin{equation}
\left\langle p_{kq}\right\rangle 
\approx \Delta t(g_{kq}/\hbar )^2
\int K(\tau )\exp (iE_{kq}\tau /\hbar )d\tau 
\propto \Delta t\,, 
\end{equation}
Now, the fixed scale $\Delta E $ instead
of $2\pi \hbar /\Delta t $ determines energy region 
available for transfers from a given level. 

3. We are most interested in the summary jump probability. 
It can be represented as 
\begin{equation}
p_k\equiv \sum_qp_{kq}=\int \int_0^{\Delta t}
\Gamma _k(t_1-t_2)Z(t_1)Z^{*}(t_2)dt_1dt_2 \,, 
\end{equation}
with the integral kernel  
\[
\Gamma _k(\tau )=\sum_q(\frac{g_{kq}} 
\hbar )^2\exp (i\tau E_{kq}/\hbar )
\]
Analytical properties of such kernels dictated by quantum 
discreteness play a key role in the theory. In the continuous
spectrum approximation this kernel turns into a function which 
quickly (integrably) and irreversibly decays to zero, for instance, 
delta-function. But in reality this kernel is quite non-local 
and never decays (instead, sometimes it returns to a value of 
order of its value at $\tau =0 $). In principle, this property 
is nothing but reflection of the unitarity of quantum dynamics. 
If take, for visuality, the equidistant spectrum at side accepting 
jumps,  $E_{kq} =n\delta E+\varepsilon _{k} $ ,  
with integer $n$ , then  
\begin{equation}
\Gamma _k(\tau )=\frac 1{\tau _t}
\exp (i\varepsilon _k\tau /\hbar )
\sum_n\delta (\tau -n\tau _g)
\end{equation}
Now the third point is obvious: 
if $\Delta E>\delta E$  then the mean jump probability 
 $\left\langle p_{k}\right\rangle $  $\approx \Delta t/\tau _{t}$  
practically coinsides with what is used in kinetics. 
Hence, due to the noise transfers into discrete spectrum realize
as successfully as to continuous spectrum. 

But, of course, the noise can not make the ''probabilities per
unit time'' be better certain.  
The most inportant point is that due to the discreteness the 
summary jump probabilities also have non-zero fluctuations. 
With accounting for the white-noise character of $Z(t) $ , 
variancies of jump probabilities can be expressed as 
\begin{equation}
\left\langle p_{k},p_{k}\right\rangle 
\approx \int ...\int_0^{\Delta t}\Gamma _k(t_1-t_2)
\Gamma _k(t_3-t_4)K(t_1-t_4)K(t_3-t_2)dt_1...dt_4
\end{equation}
Here under condition $\Delta E>\delta E$ only 
regions $t_1\approx t_4,$ $t_3\approx t_2$ 
are significant but many delta functions from (11) contribute. 
The estimate of this integral yields 
\begin{equation}
\left\langle p_{k},p_{k} \right\rangle 
\approx \frac{\Delta t^2}{\tau _t^2\tau _g} 
\int \left| K(\tau )\right| ^2d\tau 
\approx \frac{\tau _{coh}}{\tau _g} 
\left\langle p_k\right\rangle ^2 = 
\frac{\delta E}{\Delta E} 
\left\langle p_k\right\rangle ^2
\end{equation}
(we took into account that ''width'' of delta functions determined 
by inverse energy band is wittingly smaller than $\tau _{coh}$ ).
Notice that conributions to $p_{k} (\Delta t,U) $  
from different parts of the observation time are completely  
statistically correlated although produced by independent pieces 
of realization  $u(t)$ . Such unusual transport statistics 
were considered in [4,23,24,27]. Here, it is due to 
that transport is governed by subsequent time summation of  
the amplitudes, not probabilities.

Below mutual correlation between fluctuations of probabilities 
of jumps from different levels will play an important role. 
It reads as   
\begin{equation}
\left\langle p_{k_1}, p_{k_2}\right\rangle 
\approx \left\langle p_{k_1}\right\rangle 
\left\langle p_{k_2}\right\rangle 
\frac {\delta E }{\Delta E } S(E_{k_1}-E_{k_2}) \,,  
\end{equation} 
where function 
\[
S(E)=\int \exp (iE\tau /\hbar ) \left| K(\tau ) 
\right| ^{2} d\tau \left[ \int \left| K(\tau ) 
\right| ^{2} d\tau \right] ^{-1}
\]
describes its dependence on energy
distance between two levels, $E $ .
According to these formulas, 
the ''uncertainty principle'' takes place:  
with increasing $\Delta E$ all fluctuations of jump probabilities
decrease but instead their correlations spread to more and more
wide energy distancies. The rates of charge injection from
levels belonging to the same ''coherence region'' $\Delta E $ 
are closely correlated.

At $\Delta E<\delta E$ fluctuations of jump probabilities 
may increase up to 100-percent value or even higher. In this 
extremal case quantitative estimates are sensible to details 
(commensurability etc.) of energy spectra, therefore one needs
in some statistics of energy levels. Besides, in this case the
noise induced renormalization of mean jump probabilities and
consequently average transport current and CVC becomes significant. 
All the above results can be extended to when
inequality (5) is not satisfied. 

\section{Conductance fluctuations}

1. Consider fluctuations of the charge transported through the
junction between outward leads. We will confine ourselves by the
case of not too low temperature, $T>>\delta E $ , and not too
high external voltage, $U<T/e$ . 

Below let $\Delta Q $ denote the random value.
It consists of two parts,   
\[
\Delta Q =\Delta Q_{th} +\Delta Q_{ex} 
\]
Here first term is contribution of fast thermal (shot) current 
noise caused by uncertainty of jump instants. This part contributes
to variance of $\Delta Q$ even in equilibrium (at $U=0$ ) and can
be estimated with the help of fluctuation-dissipation theorem
(Nyquist formula),  
\[
\left\langle \Delta Q^{2}_{th} \right\rangle
\approx 2TG\Delta t 
\]
We are more interested in the second term which besides mean value of 
transport current includes also excess transport fluctuations caused
by the above discussed uncertainty of jump probabilities. 

Within statistical language,  $\Delta Q_{ex} $ may be 
defined as conditional average value of $\Delta Q $ under 
fixed $p_{kq} $ . It is clear, from simple thermodynamical 
reasonings, that this conditional average has the same sign 
as $U $ and disappears at $U=0 $ . Therefore, it can be 
represented as the sum of definitely directed jumps (for instance, 
from left to right at $U>0 $ ). This circumstance allows us to 
concentrate at the excess contribution without special analysing 
correlations between oppositely directed electron transfers. 
These correlations create no obstacles to arbitrary low-frequency
transport fluctuations (distributed at times longer $\tau _{c}$ )
since such fluctuations do not touch inner statistical balance
in the junction [26,27]. Thus we can assume that all the
correlations are already included into statistics of voltage
noise $u(t)$ . All the more this is reasonable in view of the
appointed fact that details of the noise statistics are unimportant.

2. Therefore, let us attribute formulas of previous Section to
the probabilities of one-directed excess jumps responsible for 
transport. For a first, one may neglect correlations between random
energy positions of ''active'' levels which inject the excess jumps,
and preliminarily perfom the averaging with respect to the postions. 
Then the start expression for the excess transport looks as
\begin{equation}
\Delta Q_{ex}=e\sum_k[f(E_k)-f(E_k+eU)]p_k(\Delta t,U)\,,
\end{equation}
with $f(E) $ being the Fermi distribution function.  
Below with no noticable losses we may 
put on  $p_{k}(\Delta t,U)\approx $  $p_{k}(\Delta t, 0) $ . 
If one additionally neglected the randomness of probabbilities, 
then Eq.15 would reduce to the well known expression for mean
transport current. It must be underlined that Eq.15 
completely accounts for the Fermi statistics in both electrodes.
The result of averaging (15) coinsides with Eq.1. 

Consider fluctuations of the
conductance $G=\Delta Q_{ex}/U\Delta t $ implied by Eq.15,
omitting details of calculations.      
At small coherence time, what corresponds to ''large'' junction,
dense levels, and  $\Delta E>\delta E$ , with the help of formulas
of previous section the variance of (15) can be transformed into 
\begin{equation}
\left\langle \Delta Q_{ex},\Delta Q_{ex} \right\rangle
\approx (GU\Delta t)^2\frac{\delta E}{\Delta E}
\int W(E^{\prime })W(E^{\prime 
\prime })S(E^{\prime }-E^{\prime 
\prime })dE^{\prime }dE^{\prime \prime }
\end{equation}
Here function   
\[
W(E)=[f(E)-f(E+eU)]\,/eU \approx -\partial f(E)/\partial E 
\]
has the sense of the "one-particle" probability density
distribution of energy of active levels.  
The Eq.16 yields rather universal estimate 
\begin{equation}
\delta G^{2} \equiv 
\frac{\left\langle G,G\right\rangle }{
\left\langle G\right\rangle ^2} 
\approx \frac{\delta E}T\,,
\,\,\,(\tau _{coh}<\tau _g)
\end{equation}

At large coherence time, what corresponds to ''small'' junction, 
rarefied levels, $\Delta E<\delta E$ , and weakly correlated
jumps, i.e. to the above mentioned extremal situation, so
definite estimate is impossible, because conductance fluctuations
are very sensible to statistics of energy levels, first of all
to their commensurability. In this case ''everything is possible'', 
\begin{equation}
\frac{\delta E}T< \delta G^{2} <1\,,
\,\,\,(\tau _{coh}>\tau _g)\,,
\end{equation}
up to relative fluctuations of order of unit. 

3. As it is seen, in general the discreteness directly serves 
as the measure of conductance fluctuations. At the same time, 
factor  $\Delta E $ which characterizes intensity of  
''enviromental noise'' does not contribute to (17). 
This is likely if the mean number of active 
levels, $N \equiv eU/\delta E $ , is not large and these levels 
make take arbitrary relative positions, in particular, 
all occur in one and the same coherence region  $\Delta E $ . 
But the latter case is impossible when  $N $ exceeds total 
number of levels in such a region, $\Delta E/\delta E $ . 
Consequently, at  $eU>\Delta E $  permissible energy 
distributions of active levels are more uniform, therefore 
jumps from them are less correlated. Under increase of applied 
voltage, this may result in supressing conductance fluctuations 
even at $eU<<T $ when average conductance is still more 
or less constant. 

This effect is not scoped by the approximation (15). Formally,
as the Eq.16 shows, the matter is that it corresponds to the 
factored pair ("two-particle") energy 
distribution  $W(E^{\prime })W(E^{\prime \prime }) $ .  
But it is not hard to improve our analysis. Let us enumerate 
active levels in the increasing energy order. The typical
distance between two levels with numbers  $j>i $ can not be 
smaller than  $\approx (j-i)\delta E $ . Therefore, any 
pair of numbers implies its own pair distribution,  
\begin{equation}
W_{ij}(E^{\prime },E^{\prime \prime })
\approx W(E^{\prime })W(E^{\prime
\prime })\vartheta (\left| E^{\prime 
\prime }-E^{\prime }\right| -\left| j-i\right| \delta E) 
\end{equation} 
wher $\vartheta ()$ is step-like function. 
Further, instead of (15), we should write  
\begin{equation}
\Delta Q_{ex}=e\sum_{j=1}^n p_{k_j}(\Delta t,U)\,\,, 
\end{equation}
thus taking into consideration all possible random positions 
of active levels at energy scale (i.e. substituting average  
occupancies by factual unit). Here  $n $ is random total number
of active levels whose mean value equals to $N $ .  
Then calculate variance of this expression. Since at $N\sim 1 $ 
the answer must coinside with previous result, 
we may put on  $N>>1$ , $n=N $ . 
Performing the averaging with respect to both random energy 
positions and jump probabilities yields 
\begin{equation}
\left\langle \Delta Q_{ex},\Delta Q_{ex} \right\rangle 
\approx e^2(\gamma \Delta t )^2 \frac {\delta E}{
\Delta E} \sum _{i,j=1}^{N} \int W_{ij}(E^{
\prime},E^{\prime \prime })S(E^{\prime }-E^{\prime 
\prime })dE^{\prime}dE^{\prime \prime }
\end{equation}
After elementary manipulations we obtain instead of (17) 
the corrected estimate of conductance fluctuations, 
\begin{equation}
\delta G^{2} \approx \frac {\delta E}{T} D(eU) \,\,\,, 
\end{equation}
where function  
\begin{equation}
D(X)=\frac {1}{X}\int_0^XdE\int_E^\infty S(E^{\prime })dE^{\prime }
\cdot \left[\int_0^\infty S(E)dE\right] ^{-1}
\end{equation}
describes their dependence on external voltage. 
For example, in the case of exponential phase correlation we have
\[
K(\tau )=\exp (-\left |\tau \right |/\tau _{coh})\,,
\,\,D(X)=\frac {1}{X} \int ^{X}_{0}
\left (1-\frac {2}{\pi }\arctan \frac {E}{\Delta E}\right )dE
\]
The esimate (18) modifies in analogous fashion. 

According to these formulas, at  $eU>\Delta E $ the fluctuations 
decrease approximately as $\propto U^{-1} $ . One can say, that
the increase of applied voltage results in increase of effective
number of statistically independent (energetical) tunneling
channels, and relative conductance fluctuations decrease
inversely proportionally to the number of channels. 
Thus the degree of discreteness,  $\delta E $ , as well as 
the degree of voltage noise, $\Delta E $ , both have direct 
reflection in ''current-noise characteristics'' if not in 
current-voltage characteristics. The clear separation of 
duties of these two parameters gives hope that their 
significance will be maintained in a more rigorous theory. 
 
4. What is for the transparency of junction, in relative units 
it slips out from the estimates. In this sense it can not 
be treated as ''small expansion parameter'', as well as 
in the sense of relation between excess transport noise and
thermal (shot) noise. Though the shot noise contribution to
variance of the transported charge is proportional to the
transparency, while excess contribution is proportional to
it squared, the latter grows as the observation time squared, 
while the former grows linearly with time. Therefore, excess
transport noise inevitably dominates at large times and low
frequencies. It is easy to verify that at $eU\sim T $ this takes
place already after a time of order of one jump
duration, $\tau _t $ . In fact, in experiments and practice excess
noise noticably dominates at low frequencies at any realistic
operational currents. Besides, naively formal reasonings on
''powers of transparency'' are wrong in view of that  
the random phase $\varphi (t) $ involves in some complicated 
non-perturbative way all orders of the transparency. 

\section{Discussion and comparison with experiments}

1. Experimental confirmation of the theory can be found in the 
work [41] devoted to 1/f noise in films of the cermet 
(nano-composite) formed by $Ni$ nano-particles in matrix $Al_2O_3$ . 
In this system the parameters of a typical elementary tunnel 
junction between neighbouring particles were 
 $\delta E\approx 0.2\,$meV , 
 $d\approx 2\,$nm, $C\approx 5\cdot 10^{-6}\,$cm, 
 $E_C\sim T$ (at room temperature), 
 $R\approx 30\,$MOhm, what means  
 $\tau _{g}\approx 3\cdot 10^{-11}\,$s, 
 $\tau _{c}\approx 1.5\cdot 10^{-10}\,$s 
 and $\tau _{t}\approx 3\cdot 10^{-8}\, $s. 
In this system very intensive conductance fluctuastions
were observed with relative spectral
power  $S_{\delta G}(f)\approx \alpha /N_gf\, $, 
where  $\alpha \approx 6\cdot 10^{-3}\,$ , and $N_g\,$ is number 
of metal granules in the sample. Since $E_c\sim T\,$ , 
then  $N_g\,$ approximately coinsides with the number of movable 
(simultaneously transported) electrons in the sample [41], 
therefore it was almost ''standard'' 1/f-noise with ''classical'' 
Hooge constant [1-4].  

These data correspond to the
noise  $S_{\delta G}(f)\approx \alpha /f\,$  
in a separate elementary junction. Since both the inequalities
(2) and (5) are well satisfied, we may assume that this noise is
due to the mechanism under consideration. For stationary 1/f noise,
the above calculated conductance variance related to a definite
smoothing time can be connected with 1/f spectrum as
\[
\delta G^2 \approx fS_{\delta G}(f) \ln (\Delta t /\tau_{c}) 
\approx \alpha \ln (\Delta t /\tau_{c}) 
\]
At $\Delta t \sim \tau_t $ this gives the value $\sim 0.03 $ .
Formula (17) at  $\delta E = 0.2\,$meV and room temperature
yields $\approx 0.008 $ . This is good agreement, in view of
that strictly speaking this case must be subjected to formula
(18). Indeed, due to above estimates, 
here  $\tau_{coh} \sim \tau _{c} > \tau_{g} $ , 
hence  $\Delta E $ is of order of $\delta E $ or even smaller
what corresponds to the extremal situation.    

The remarkable observation of [41] is 1/f-noise sensibility to  
the discreteness of electron energy spectra in metal granules. 
When applied voltage per elementary junction 
exceeds $\delta E/e\,$ , i.e. when more than one level in 
a granule is active, relative 1/f noise intensity decreases 
as inverse voltage, although average CVC   
is still Ohmic up to $\sim T/\delta E\sim 100\,$ times larger
voltages. In view of  $\Delta E \sim \delta E $ , it is clear
that in respect to this effect our theory is in full agreement 
with the experiment. 

2. In case of cermet the discreteness of levels is directly 
determined by a volume of junction sides, i.e. metal nano-particles. 
Obviously, in a junction with massive sides (electrodes) 
the quantity  $\delta E $ also must be determined by volume 
of the space region which is physically available for electron
jumps, that is by geometry of junction and processes of electron 
interactions and scatterings in electrodes. 
If temperature is not too low then one has no many variants. 
The available region must be restricted by junction area, $A $ , 
and non-elastic free path in sides,  $\lambda $ . 
In other words, this is the region in which level spacing 
has the order of broadening of levels due to non-elastic relaxation. 
Of course, now it would be more precise to say not about literally
levels but about statistics of energy repulsion of spatially close
electron states [40], which has a sense and acts also at non-zero 
temperatures [39]. Thus the term discreteness serves as synonym 
of uncertainty: nuances of electron states depend on the
''enviromental noise'' and can not be controlled with precision 
better than  $\delta E $ .  

3. Consequently, in case of massive metallic electrodes one 
can estimate  $\delta E\sim E_{F}a^3 /A\lambda $ ,  
with $E_F $ being Fermi energy and $a $ atomic size. 
Then, if roughly relate  $\lambda $ to conductivity  
in electrodes, $ \sigma \sim \lambda/a^{2}R_{0} $ , 
the estimate (17) transforms into 
\begin{equation}
\delta G^{2} \sim 
\frac {E_F}{T} \cdot \frac {a^2}{A} \cdot 
\frac {\sigma _{min}}{\sigma }
\end{equation}
where $\sigma _{min} \sim (aR_0)^{-1} $ is minimal 
metallic conductivity. In particular, let the metal be so clear
that the electron-phonon mechanism of relaxation is dominating. 
Then, according to [43], 
\[
\sigma \sim (\hbar e^{2}n_{e}/m^{*}T)(T_{D}/T)^4 
\]
(all notations are standard), and one may expect that at 
temperatures lower than Debay temperature conductance fluctuations 
are proportional to $T^4 $ . 

If say about possible orders of value only, then at
standard $E _F $ the Eq.24 yields  
\begin{equation}
\delta G^{2} \sim (a^{2}/A)(T/T _{D} )^4 
\end{equation}
From the other hand, left side of (25) may be connected, 
like it was made above, with spectral power of 1/f-noise. 
Then in the case of micro-junction with area $10^{-9}\,cm^2 $ 
investigated in [37], under assumption $T \sim T_{D} $  
we obtain the estimate $ fS_{\delta G}(f) \sim 10^{-7} $ . 
This value is in agrtement with meausurements in 
[37] at 260 K . Besides, the estimate (25) naturally explain  
sharp (approximately by two orders of magnitude) increase
of 1/f-noise under increasing temperature from 100 K to 300 K 
as was observed in [37]. To explain this fact the authors 
supposed sharp step in the activation energy distribution 
of fluctuators (although this assumption contradicts to 
uniformity of the distribution wished for 
composing spectrum 1/f). From our look, likely in [37] two 
types of low-frequency noise were observed: one has structural 
origin and dominates below 100 K, while another is just what 
was considered and dominates at higher temperatures. 

4. In this connection, it is useful to notice that generally 
summation of structural fluctuations and fundamental flicker 
noise under our attention may not obey trivial adding rules [4]. 
Clear 1/f spectrum corresponds to fast forgetting the past [4,27], 
while entangling between the fundamental noise and actually slow 
relaxation processes (for instance, heat transport) may lead to 
somehow deformed spectrum (for instanse, $1/f^{\gamma }$ 
with $\gamma >1$ )[4]. In particular, structural defects can
affect 1/f noise although not being its origin. For example, 
elastic electron scattering by impurities supresses mobility 1/f 
noise [2]. Experiments with metallic films show that 1/f noise
there is strongly influenced by vacancies [42]. But this effect
can not be explained in terms of thermally activated 
diffusion of vacancies. Indeed, unlikely separate diffusive 
jumps of vacancies have very large variance of activation 
energies. What is for the diffusion as the whole it is described
by the same equations as temperature diffusion, but the latter
far ago was removed as hypothetical source of 1/f spectrum [1-4].
Perhaply, vacancies promote non-elastic scattering of electrons 
and in this way become mediator for 
the ''1/f-noise from lossing memory''. 
 
\section{Conclusion} 

As it was demonstrated, the careful treatment of quantum discreteness 
displays that amplitudes of elementary quantum transfers being 
sensitive to surrounding noise of the whole system behave like random 
Brownian walks what leads eventually to low-frequency flicker 
fluctuations of transport rates. These fluctuations have no
characteristic time scale and are indifferent to duration 
of observation and time averaging. 

Formally, in our approximation restricted by time 
intervals $\Delta t\sim \tau _{t}$ we obtained fluctuations 
with perfectly undecaying correlations corresponding to 
spectrum  $\propto \delta (f) $ . This resembles static universal 
conductance fluctuations [40], although by essence thats are quite 
different things: one starts from inelastic relaxation and 
decoherence while another finishes at them. We expect that rigorous
many-particle theory will extend our results (and first of all the
scaleless character of conductance fluctuations) to arbitrary long
time intervals. In this theory quantities kindred to above $p _{k}\,$ 
will be measure of number of electron jumps but will remain 
proportional to both time and transparency (highest orders of 
transparency will be absorbed by the random phase, - see notion at 
end of Sec.3), while spectrum  $\propto \delta (f) $ must transform 
to a spectrum  $\propto 1/f $ with the same frequency dimensionality. 
Formal grounds of such a confidence will be considered elsewhere. 
Its principal motivation is that the only source of our results
is the trivial quantum mechanical rule that even at presence of 
noise and decoherence any kinetics is governed by summation or 
other play of amplitudes, not of intermediate probabilities.  

\,\,

\,\,

{\bf REFERENCES}

\,\,

1. P.Dutta, P.Horn, Rev.Mod.Phys., 53(3), 497 (1981).

2. F.N.Hooge, T.G.M.Kleinpenning, L.K.J.Vandamme, 
Rep.Prog.Phys., 44, 481 (1981).

3. M.B.Weissman, Rev.Mod.Phys., 60, 537 (1988).

4. G.N.Bochkov, Yu.E.Kuzovlev, Sov.Phys.-Usp., 26, 829 (1983).

5. M.B.Weissman, Rev.Mod.Phys., 65, 829 (1993).

6. B.Raquet, J.M.D.Coey, S.Wirth and S. Von Molnar, 
Phys.Rev., B59, 12435 (1999).

7. J.G.Massey and M.Lee, Phys.Rev.Lett., 79, 3986 (1997).

8. M.J.C. van den Homberg, A.H.Verbruggen et al., 
Phys.Rev., B 57, 53 (1998).

9. A.Ghosh, A.K.Raychaudhuri et al., Phys.Rev., B 58, R14666 (1998).

10 X.Y.Chen, P.M.Koenrad, F.N.Hooge et al., 
Phys.Rev., B 55, 5290 (1997).

11. G.M.Khera and J.Kakalios, Phys.Rev., B 56, 1918 (1997).

12. M.Gunes, R.E.Johanson and S.O.Kasap, 
Phys.Rev., B 60, 1477 (1999).

13. K.M.Abkemeier, Phys.Rev., B 55, 7005 (1997).

14. G.Snyder, M.B.Weisman and H.T.Hardner, 
Phys.Rev., B 56, 9205 (1997).

15. V.I.Kozub, Solid State Commun., 97, 843 (1996).

16. A.Lisauskas, S.I.Khartsev and A.M.Grishin, 
J.Low.Temp.Phys., MOS-99 Proceedings.

17. A.Lisauskas, S.I.Khartsev, A.M.Grishin and V.Palenskis,
Mat.Res.Soc.Proc., Spring-99 Meeting.

18. V.Podzorov, M.Uehara, M.E.Gershenson and S.-W.Cheong, 
lanl arXiv cond-mat/9912064.

19. M.Viret, L.Ranno and J.M.D.Coey, Phys.Rev., B 55, 8067 (1997).

20. J.M.D.Coey, M.Viret and S. von Molnar, 
Adv. Phys., 48, 167 (1999).

21. P.Bak. Self-organized criticality: why nature is complex. 
Springer, N-Y, 1996.

22. J.L.Tandon and H.P.Bilger, J.Appl.Phys., 47, 1697 (1976). 

23. Yu.E.Kuzovlev, G.N.Bochkov, Izv. VUZov. Radiofizika, 
26, 310 (1983); 27, 1151 (1984) (transl. in Radiophysics and 
Quantum Electronics, N.3, 1983, and N.9, 1984).

24. Yu.E.Kuzovlev, Sov.Phys.-JETP, 67, N.12, 2469 (1988).

25. Yu.E.Kuzovlev, JETP, 84, N.6, 1138 (1997).

26. Yu.E.Kuzovlev, Phys.Lett., A 194, 285 (1994).

27. Yu.E.Kuzovlev, lanl arXiv cond-mat/9903350.

28. N.S.Krylov. Works on the foundations of statistical physics. 
Princeton, 1979.

29. J.P.Hessling, Yu.Galperin, M.Jonson, R.I.Shekhter and 
A.M.Zagoskin, Aplied Physics Report 94-38. 

30. Yu.M.Galperin, V.G.Karpov, V.I.Kozub, 
Zh. Eksp. Teor. Fiz., 95, 1123 (1989).

31. Yu.V.Nazarov, Zh. Eksp. Teor. Fiz., 95, 975 (1989).

32. M.H.Devoret, D.Esteve, H.Grabert et al., 
Phys.Rev.Lett., 64, 1824 (1990).

33. S.M.Girvin, L.I.Glazman, M.Jonson et al., 
Phys.Rev.Lett., 64, 3183 (1990).

34. I.G.Lang, Yu.A.Firsov, Zh. Eksp. Teor. Fiz., 43, 1843 (1962).

35. Yu.A.Genenko and Yu.M.Ivanchenko, Phys.Lett., 126, 201 (1987).

36. Yu.M.Ivanchenko, Yu.V.Medevedev, Fiz. Nizk. Temp., 2, 142 (1976).

37. C.T.Rogers and R.A.Buhrman, Phys.Rev.Lett., 53, 1272 (1984).

38. Yu.A.Genenko, Yu.M.Ivanchenko, Teor. Mat. Fiz., 69, 142 (1986).

39. G.Casati and B.Chirikov. Fluctuations in quantum chaos. 
Preprint, Budker Inst. of Nuclear Physics SB RAS, 1993.

40. C.W.J.Beenakker, Rev.Mod.Phys., 69, N 3, 731 (1997).

41. J.V.Mantese, W.I.Goldburg, D.H.Darling et al., 
Solid State Commun., 37, 353 (1981).

42. G.P.Zhigalski, Usp. Fiz. Nauk, 167, 623 (1997). 

43. E.M.Lifshitz, L.P.Pitaevski. Physical kinetics. Moscow, 1974.

44. A. van Oudenaarden, M.H.Devoret et al., 
Phys.Rev.Lett., 78, 3539 (1997).

45. Yu.V.Nazarov, Sov.Phys.-JETP, 71, 171 (1990). 

46. Yu.E.Kuzovlev, lanl arXiv cond-mat/0004398. 

\end{document}